# Dummy variables and their interactions in regression analysis: examples from research on body mass index

Manfred Te Grotenhuis

Paula Thijs

**Abstract**

This paper is especially written for students and demonstrates the correct use of nominal and ordinal scaled variables in regression analysis by means of so-called 'dummy variables'. We start with examples of body mass index (BMI) differences between males and females, and between low, middle, and high educated people. We extend our examples with several explanatory (dummy) variables and the interactions between dummy variables. Readers learn how to use dummy variables and their interactions and how to interpret the statistical results. We included data, syntax (both SPSS and R), and additional information on a website that goes with this text. No mathematical knowledge is required.

## 1. Introduction

When teaching regression techniques to social science students, we experienced two subjects to be difficult for them: the use of so-called 'dummy variables' and the interactions between them. Many statistical textbooks leave these matters untouched or approach it rather technical (an exception is Melissa Hardy's *Regression with dummy variables* (1993) from which we learned a lot ourselves). That is the reason to focus here solely on 'dummy variables' and explain their use and interpretation using examples from research on body mass index (BMI). The mathematical information is kept to a minimum. This text therefore is meant for anyone who is interested in the use of dummy variables in regression models.

*The major learning outcome is that the reader will be able to use dummy variables and their interactions and to interpret the statistical results adequately.*

To achieve our goal, we will:

- Clarify the concepts of dummy variables and interaction variables in regression analysis;
- Show how dummy variables and interaction variables are used in practice;
- Provide syntax in SPSS and R for practical use.

As a leading example, we use 3 national surveys containing the body mass index (BMI) of 3,323 individuals aged between 18 and 70 in the Netherland collected in 2000, 2005, and 2011 by Rob Eisinga and associates (2002, 2012a/b). The BMI (in $kg/m^2$) is based on respondents' self-reported height and weight. It presents a somewhat underestimated figure (roughly one full point lower) of the 'true' BMI based on measured height and weight as Arno Krul and associates (2011) found. We will start with straightforward examples of BMI differences between males and females, and we will end our contribution with full-fledged regression models with several explanatory variables and their interactions. All data, results (in PDF format), syntax, and additional information are available through the Internet for free: [website](website).

## 2. Dummy variables: what are they?

In every statistical textbook you will find that in regression analysis the predictor variables (i.e., the variables that explain/predict the outcome variable) have to be interval or ratio scaled. These two types of variables have values on a fixed scale with equal distances. For instance *time* indicated by a clock is interval scaled: the interval between its subsequent values (in seconds, minutes, or hours) is always the same. Ratio scaled variables have fixed intervals too, but on top they contain a meaningful and absolute zero point, such as *age*: everybody's life starts at 0 (years). In social sciences, interval/ratio scaled variables are probably outnumbered by nominal and ordinal scaled variables. Nominal variables have values as well but they serve only to distinguish between categories. Take for instance *marital status* with categories 'married', 'unmarried', 'divorced, and 'widow(er)'. You may attach values 1, 2, 3, and 4 to them, but 23, 67, 0, and 8 will do as well. In ordinal variables the values are not as unordered as in nominal variables: their values express a hierarchy like in *educational attainment*: elementary school is lowest, followed by junior high, senior high, college, and at the top we find university. To ascertain this ordering we may use 1, 2, 3, 4, 5, and 6 but 1, 4, 5, 8, 10, and 20 is as good. Because nominal and ordinal scaled variables have no nicely defined scales with fixed intervals, they are not well-suited as predictor (x) variables in regression models. To include them in these models their categories have to be transformed into so-called 'dummy' variables first. Popular is 'dummy coding' and this type of transformation will be explained here, while other transformations (or 'parameterizations') are available on the website that goes with this contribution. For ease of interpretation we will use ordinary least square (OLS) regression models in our examples, but our explanation can be generalized to any type of regression model, such as logistic regression analysis.

### 3. First example: the BMI gender gap

For many it may come as a surprise to find that the variable *sex*, with categories 'male' and 'female' is not a nominal variable. The simple reason is that it contains only two categories and this makes it formally an interval/ratio variable. This is easily seen when code 0 for males is used and code '1' for females. The average (only meaningful in interval/ratio variables) then equals the proportion of females in the data set. For instance, if the number of females and males is equal, the average is (0 + 1) / 2= .5, i.e., 50% of all respondents are female. In statistics, variables like *sex* are labeled 'dichotomous' variables. Because any variable that has only codes 0 and 1 is a ratio variable, we can include them in regression models and obtain meaningful results. In the following examples we take the variable *body mass index* (ratio scaled) as dependent variable and use the variable *female* (0 = male, 1=female) as predictor. Note that we prefer the label 'female' over 'sex', because the former indicates directly that females are coded 1. In OLS regression models this leads to the following equation:

> *BMI*$_i$ = a + b * *female* + *e*$_i$          (1)
> - Reference category: male (code 0)

Equation (1) says that the individual (i) BMI score is equal to the sum of *a*, *b* times *female,* and *e*$_i$. The *e* stands for error which is the difference between the estimated BMI (a + b * *female*) and the actual observed individual BMI score. If males have code 0, then *b* * *female* =

$b * 0 = 0$, hence their BMI is equal to $a + e_i$, and if females have code 1, their BMI equals $a + b * 1 + e_i = a + b + e_i$. In most cases researchers are not interested in individual BMI scores but in mean BMI scores per group. For this, we eliminate $e_i$ from equation (1), and get the estimated mean BMI scores in our sample:

> Mean (*BMI*) = *a* + *b* \* *female*   (2)
> - Reference category: male (code 0)

So, in our data set the sample mean BMI in males equals *a*, while the sample mean BMI in females is equal to *a* + *b*. This is true because we coded males to be 0 and females to be 1. Now suppose we want to know whether Dutch adult females on average have a *lower* BMI than Dutch adult males, like for instance Arno Krul and associates (2011) found in their study among Dutch, Italians, and Americans. The value of interest then is *b* in Equation (2), as it denotes the sample mean difference in BMI between males (*a*) and females (*a* + *b*). Because we use a random sample from the Dutch population, we like to statistically test this mean BMI difference *b*. The null hypothesis in such a test states that there is no BMI gender gap in the target population, so *b* is assumed to be 0. The research hypothesis is that females on average have a lower BMI compared to males, so *b* is hypothesized to be lower than 0. In our sample the mean BMI difference *b* amounts to -.51 (see Table 1). Next, we have to calculate the probability to find -.51 or even a more negative value in the sample under the null hypothesis of no BMI gender gap in the target population (b = 0). This probability (or *p-value*) turns out to be lower than .01 (see Table 1) which is lower than standard test criteria (α) of .05 and even lower than .01. Therefore we call the outcome of -.51 statistically significant: it is beyond reasonable doubt that the mean BMI in Dutch adult females is lower than in males in 2000-2011.

The value *a* amounts to a significant 25.23 (see Table 1), so the estimated mean BMI in males in our sample is just over 25 (in the literature overweight starts at a BMI of 25, while a BMI lager than 30 indicates obesity). We can calculate the estimated mean BMI for females from these results; they have on average a BMI of 24.72 (calculation: 25.23 + -.51). In Figure 1, we graphically show the mean BMI scores for males (25.23) and females (24.72), and the BMI gender gap (-.51).

**Table 1** The estimated BMI means for males (*a*) and the BMI gender gap (*b*)

| | coefficients | standard error | t-value | p-value (2-tailed) |
|---|---|---|---|---|
| a | 25.23 | .10 | 260.90 | <.01 |
| b *female* (0=male, 1=female) | -.51 | .13 | -3.82 | <.01 |

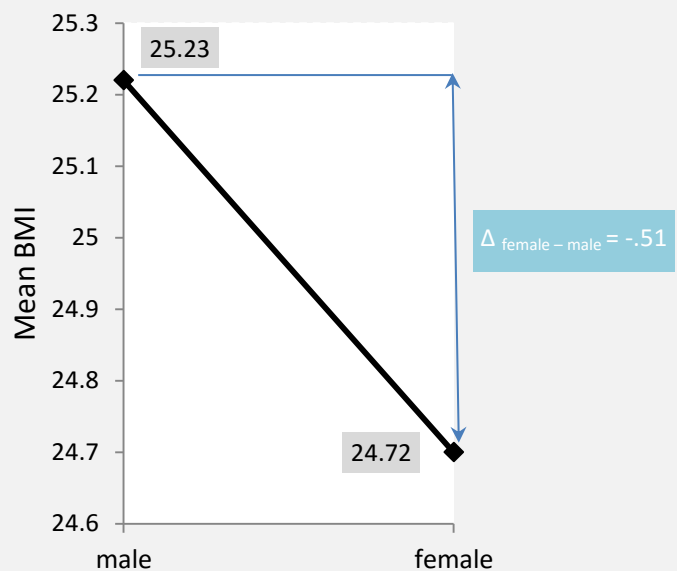

**Figure 1** The mean BMI in males and females and their mean BMI difference

## 4. Second example: the BMI differences between the low, middle, and high educated

As our first example showed, we can include 0/1 coded variables in regression models and get meaningful results. In this lies the solution for the problem of including nominal and ordinal scaled variables in regression models: *convert all their categories into dichotomous variables with a 0/1 coding*. These variables are called 'dummy' variables as they replace (or 'stand in for') the original categories. To show how it works, we take the ordinal variable *educational attainment*. For didactical reasons we use only three levels of education: low, middle, and high (in research practice the number of categories may be higher).

Following Equation (2) this leads to:

$$\text{Mean}(BMI) = a + b_1 * low + b_2 * middle + b_3 * high \qquad (3)$$

Equation (3), however, has a fundamental problem. There are four values (or 'parameters') to be estimated: $a$, $b_1$, $b_2$, and $b_3$, while we only have the mean BMI scores of three levels of education to be estimated. The solution is to leave out the parameter of one of the dummy variables (the so-called 'reference category'). In our example we left the lowest level of education out of the equation, this results in:

$$\text{Mean}(BMI) = a + b_1 * middle + b_2 * high \qquad (4)$$
- Reference category: low education

Because individuals with a low level of education have code 0 on the dummy variables *middle* and *high*, Equation (4) for this category is reduced to: mean $(BMI) = a + 0 * middle + 0 * high = a$. This means that $a$ equals the mean BMI in individuals with a low level of education. For respondents with a middle education the mean BMI equals $a + b_1 * middle$. So, for these people the estimated mean BMI is equal to the mean BMI in the lowest level $(a)$ plus $b_1$. The mean BMI in highest education is equal to the mean BMI in the lowest level $(a)$ plus $b_2$. Now, we can statistically test whether the mean BMI falls when educational level rises in the Netherlands and thereby replicate a study by Silke Hermann and associates (2011) who found exactly such effect (see Table 2 and Figure 2).

**Table 2** The estimated BMI means for low education (a) and the mean differences with middle ($b_1$) and high education ($b_2$)

|  | coefficients | standard error | t-value | p-value (2-tailed) |
|---|---|---|---|---|
| a (low) | 26.12 | .14 | 183.17 | <.01 |
| $b_1$ (middle) | -1.18 | .17 | -6.76 | <.01 |
| $b_2$ (high) | -1.83 | .18 | -10.19 | <.01 |

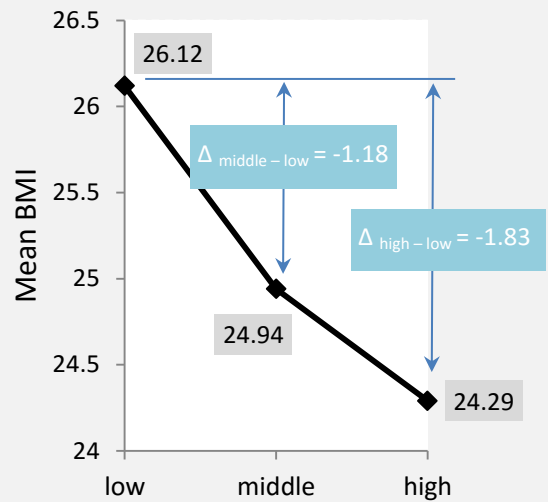

**Figure 2** The mean BMI in low, middle, and high education and the mean BMI differences between middle and low education and between high and low education

Table 2 and Figure 2 show that the low educated (the reference category) in our sample have on average an estimated BMI of 26.12. The mean BMI of the middle educated lies 1.18 points lower, which is 24.94. The highly educated have on average the lowest BMI: 24.29 and that is 1.83 points lower than the mean BMI in the low educated. These differences (-1.18 and -1.83) both are significant (p-values below .01). For a complete picture, the difference between the high and middle educated (which is 24.29 – 24.94 = -.65) must be statistically tested as well. To do this, the parameter for low educated is included in the model, while the parameter for middle education is excluded:

Mean ($BMI$) = $a + b_1 * low + b_2 * high$  (5)
- Reference category: middle education

In Equation (5) the parameter $a$ equals the estimated mean BMI among the middle educated while $b_1$ and $b_2$ denote the mean differences with the low and highly educated (see Table 3 for results and note that the mean BMI difference of -.65 indeed is significant).

**Table 3** The estimated BMI means for middle education (a) and the mean differences with low ($b_1$) and high education ($b_2$)

|  | coefficients | standard error | t-value | p-value (2-tailed) |
|---|---|---|---|---|
| a (middle) | 24.94 | .10 | 249.58 | <.01 |
| $b_1$ (low) | 1.18 | .17 | -6.76 | <.01 |
| $b_1$ (high) | -.65 | .15 | 4.41 | <.01 |

To summarize: when a researcher wants to include nominal and/or ordinal scaled variables in a regression model, then the variables' categories have to be converted into dummy variables with 0/1 coding. Generally, code 1 stands for 'this unit belongs to category *x*' and 0 stands for 'this unit does not belong to category *x*'. One can include k – 1 dummy variables, where k stands for the total number of categories in the ordinal/nominal variable. The category that is left out of the equation is called 'the reference category'. All the parameters of the dummy variables included denote the difference/deviation from this reference category. In the next section we will show that this also holds for more complex models.

## 5. Third example: multiple regression model with sex, age, education, and number of children

In this example we present a more complex model in which we test whether: a) females on average have a lower BMI than men; b) the middle and higher educated have a lower BMI than the low educated; c) the average BMI has risen since the year 2000; d) the mean BMI rises when someone has children and e) BMI on average rises with age. We simultaneously estimate all parameters because we then get the 'controlled' effects. In our second example we found that the mean BMI decreases as the level of education is higher. Further, it is still a fact that females on average have a lower level of education compared to men. If we combine these two realities, the BMI gender gap of -.51 in our first example is an underestimation. If we want to fairly compare females' and males' BMI, we must rule out the educational gender differences first. In a similar vein, we like to rule out educational differences when comparing people who have children with childless people because the latter on average are higher educated. The variable *number of children* is a ratio scaled variable but we will use dummy variables for its categories. There are two reasons for this. Firstly, this variable is very skewed: the number of respondents drops dramatically after three children (only 4% of our respondents have four children, while only one respondent has 10 children). Secondly, Debra Umberson and associates (2011) showed that the sharpest BMI increase occurs after people had their first child. This finding would be obscured when using the ratio scaled variable 'number of children' as it would (falsely) estimate a positive linear effect (i.e., a constant increase of BMI with every extra child). The equation is as follows:

Mean (*BMI*) = $a + b_1$ * *female* + $b_2$ * *middle* + $b_3$ * *high* + $b_4$ * *one child* + $b_5$ * *two children* + $b_6$ * *three children* + $b_7$ * *four or more children* + $b_8$ * *2005* + $b_9$ * *2011* + $b_{10}$ * *age*(log)   (6)

The outcomes of Equation (6) are presented in Table 4. As expected, the BMI gender gap is somewhat wider now after taking into account other relevant factors, of which education is the most important. After controlling, females have an estimated mean BMI that is -.58 points (-.51 in the first example) lower than males. As in our second example, the middle and higher educated have mean BMI's that are below that of the lower educated: -.70 and -1.42 respectively.

Respondents with one child have on average a mean BMI that is almost 1 full point higher (.97) compared to the childless, after taking into account the relevant other factors. Interestingly, this difference does not grow larger with more children. We like to note that for statistically testing the BMI difference between for instance respondents with two and three children, it is required to change the reference category. This means including the dummy for 'no children' in Equation (6) and excluding the dummy for 'two children' to have the correct reference category.

In line with global trends (see the famous worldwide overview provided by Gretchen Stevens and associates, 2012), the mean BMI has risen over the years in the Netherlands. Compared to the year 2000, the mean BMI was .21 points higher in 2005 and in 2011 it was even .40 points higher. Note that the rise in 2005 is not significant as the p-value is .16. However, this p-value is from a two-tailed test (the research hypothesis being that the BMI in 2005 is *different* from that in 2000). We, however, hold the hypothesis that the BMI has *risen* over the years, so a p-value of .16/2=.08 must be used, which makes it a significant outcome when the criteria (α) of .10 is used. Note that in many software packages, including SPSS and R, p-values are often presented from a two-tailed test.

Finally, we included age to test whether the mean BMI rises as respondents are older. Because the relationship between age and BMI is expected to be non-linear, we used the natural logarithm of age. The b-estimate for this log-transformed age variable is positive and significant, so BMI indeed seems to come with age. This increase flattens off slightly as one is older (not shown here). For interested readers we refer to our website for a graphical representation of this relationship, the syntax to create such a graph, and some additional analyses.

**Table 4** A multiple regression model for BMI including sex, education, year, number of children, and age

|  | coefficients | standard error | t-value | p-value (2-tailed) |
|---|---|---|---|---|
| *a (constant)* | 23.67 | .26 | 92.19 | <.01 |
| female | -.58 | .13 | -4.48 | <.01 |
| low | reference | | | |
| middle | -.70 | .17 | -4.07 | <.01 |
| high | -1.42 | .18 | -7.99 | <.01 |
| no children | reference | | | |
| 1 child | .97 | .22 | 4.48 | <.01 |
| 2 children | .64 | .18 | 3.55 | <.01 |
| 3 children | .83 | .22 | 3.71 | <.01 |
| 4 children or more | .90 | .31 | 2.90 | .01 |
| 2000 | reference | | | |
| 2005 | .21 | .15 | 1.39 | .16 |
| 2011 | .40 | .17 | 2.39 | .02 |
| age (log) | 1.95 | .22 | 8.74 | <.01 |

We end this example with some remarks. The value for *a* (also named 'constant' or 'intercept' in the literature) takes the value 23.67. This is the estimated mean BMI with all variables in the regression model fixed at value 0. In our case this is the estimated mean BMI for males, with a low education, interviewed in 2000, with no children and at age 18. The first four conditions are logical as they relate directly to all the reference categories that have been used (with code 0 on all dummy variables included in Equation (6)). The fifth condition (age=18), however, needs some additional explanation. Normally, the condition for age would have been the natural logarithm of 1 (which is 0). Then we would have had the mean BMI for one year old people. Our data set starts with 18 year old respondents, so that would have made no sense. Because the natural logarithm of 18 equals 2.89, we subtracted this number from the log-transformed age variable. So after subtraction, the new value 0 corresponds to 2.89 in the variable *age*(log), which is 18 in the original variable *age*. We like to note that this subtraction (or 'centering') only affects the value for *a*, and not the other effects.

A last remark is on the interpretation of the parameters for all the dummy variables in Table 4. They *always* reflect the deviation from the reference category of the original variable after controlling for relevant other variables. So the estimate for females (-.58 in Table 4) is the controlled deviation from males and males only. The estimates for the middle and high educated are controlled deviations from the low educated, and the estimates for the *number of children* all relate to controlled deviations from the mean BMI in the childless. We explicitly state this interpretation here because many (students) are led to wrongly believe that the estimates of dummy variables denote controlled deviations from the *combination* of all reference categories.

## 6. Fourth example: do educational effects differ between males and females?

Thus far we assumed that the effects for the dummy variables are equal for everyone. However, there are reasons to believe that this is not the case. For instance, the effect of education may be different for females than for males as Silke Hermann and associates (2011) found in their study. When effects differ across groups or categories we call this *interaction* or *moderation*. To explain how interaction works in regression models, we first present results from regression analyses for male and females separately (see Table 5). In this table, we present the outcomes of two separate regression analyses; one for all males in our sample, and the other for all females. As a result, we get estimates for the effects of education for males and females separately. This way, we can easily show differences in effects. Later in this section we will combine the results from Table 5 in one regression model. Table 5 therefore serves mainly as a didactical tool to show the interactions.

**Table 5** The estimated BMI means for low education (a) and the mean differences with middle and high education for males (upper panel) and for females (lower panel)

|  | coefficients | standard error | t-value | p-value (2-tailed) |
|---|---|---|---|---|
| **MALES** | | | | |
| *a (constant)* | 26.07 | .18 | 145.38 | .00 |
| low | reference | | | |
| middle | -.82 | .22 | -3.65 | .00 |
| high | -1.37 | .23 | -6.09 | .00 |
| **FEMALES** | | | | |
| *a (constant)* | 26.16 | .22 | 120.25 | .00 |
| low | reference | | | |
| middle | -1.47 | .26 | -5.59 | .00 |
| high | -2.29 | .28 | -8.32 | .00 |

In Table 5 both females and males with a low education seem to have by and large the same mean BMI (26.16 vs. 26.07). Interestingly, the BMI difference is much larger among middle and high educated respondents. According to Table 5, middle educated males have a mean BMI that is -.82 points below that of low educated males, so their mean BMI = 25.25 (26.07 + -.82). For middle educated females the mean BMI is lower: 26.16 + -1.47 = 24.69. In the high educated the mean male BMI is 24.70 (26.07 + -1.37), and for females it is 23.87 (26.16 + -2.29). We have graphed these 6 BMI means in the left panel of Figure 3, to show that the BMI gender gap widens as the level of education increases.

    The one thing that is missing is a test that tells us whether this widening BMI gender gap is significant or not. In our data set low educated females have just a little higher mean BMI than males, so their gap is close to zero (26.16 – 26.07 = .09). For the middle educated the gap is wider and reversed: females have a mean BMI that is -.56 *below* the males' mean BMI (24.69 – 25.25). For the high educated the reversed gap is even wider: -.83 (23.87 – 24.70), see the right panel of Figure 3 where we explicitly focus on these gender gaps (note that both panels in Figure 3 are equal, the only difference lies in focus).

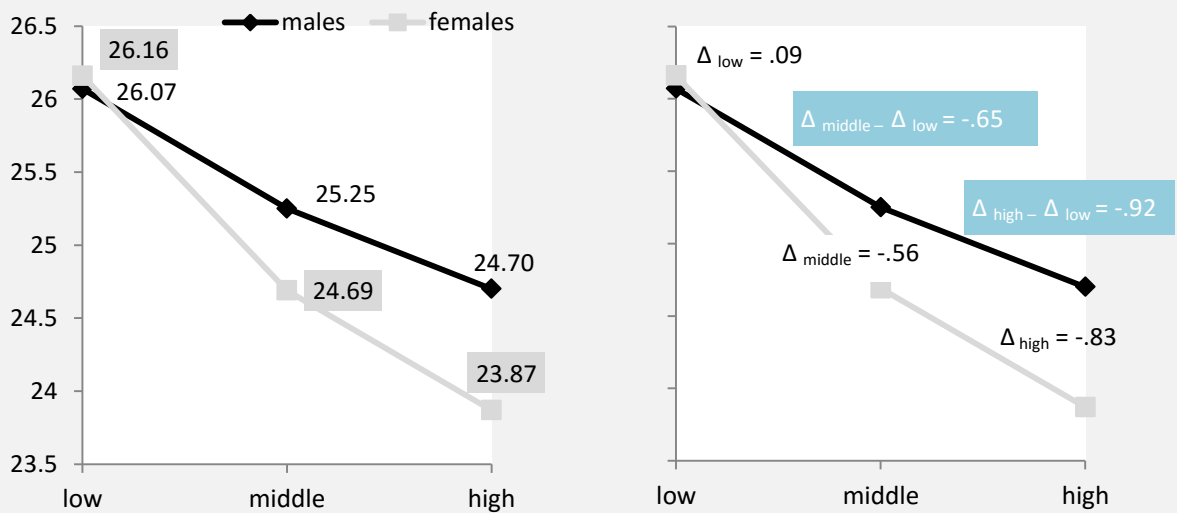

**Figure 3** The mean BMI in low, middle, and high education for males and females (left panel) and the mean BMI gender gaps per level of education + gender gap differences (right panel)

We like to test whether the BMI gender gaps of -.56 (the difference between middle educated males and females) and -.83 (the difference between highly educated males and females) are significantly reversed and wider compared to .09 (the difference between low educated males and females). Equation (7) entails those two tests:

$$\text{Mean }(BMI) = a + b_1 * female + b_2 * middle + b_3 * high$$
$$+ b_4 * female * middle + b_5 * female * high \quad (7)$$
- Reference categories: male and low education

In Equation (7) the first line is a combination of our first and second example. The second line holds the so-called 'interactions' and consist of multiplications of the dummy variable *female* with *middle* and *high*.

To read this equation, it is helpful to fill in the codes. In the dummy variable *female* the code 0 stands for males while for females the code is 1. Given these codes, for males Equation (7) is reduced to: mean $(BMI) = a + b_1 * 0 + b_2 * middle + b_3 * high + b_4 * 0 * middle + b_5 * 0 * high$. This results in: mean $(BMI) = a + b_2 * middle + b_3 * high$. Note that this reduced equation is equal to Equation (4) in our second example, only this time it is valid for males only. So, parameter $a$ is the mean BMI in low educated males, while $b_2$ and $b_3$ are the mean BMI differences between the low educated males and the middle ($b_2$) and highly ($b_3$) educated males.

For females the equation is: *mean (BMI) = $a + b_1 * 1 + b_2 * middle + b_3 * high + b_4 * 1 * middle + b_5 * 1 * high = a + b_1 + b_2 * middle + b_3 * high + b_4 * middle + b_5 * high$.* Following this equation, for low educated females the estimated mean BMI is not $a$ (for low educated males) but $a$ plus $b_1$ (calculation: $a + b_1 + b_2 * 0 + b_3 * 0 + b_4 * 0 + b_5 * 0$). So, $b_1$ is the difference in mean BMI between low educated females and low educated males: this is the first BMI gender gap (.09), see Figure 3, right panel.

For middle educated males (code 1 for *middle*, all others 0) the mean BMI amounts to $a + b_2$ (calculation: $a + b_1 * 0 + b_2 * 1 + b_3 * 0 + b_4 * 0 * 1 + b_5 * 0 * 0$). For middle educated females (code 1 for both *female* and *middle*) the estimated mean BMI is $(a + b_2) + (b_1 + b_4)$ (calculation: $a + b_1 * 1 + b_2 * 1 + b_3 * 0 + b_4 * 1 * 1 + b_5 * 1 * 0$). The BMI gender gap then is $b_1 + b_4$ (calculation: $(a + b_2) - (a + b_2) + (b_1 + b_4)$). So, the BMI gender gap for the middle educated (-.56 in Figure 3) is the sum of the BMI gender gap among the low educated ($b_1$), (.09 in Figure 3) and $b_4$. To find the value $b_4$ we have to subtract .09 from -.56, which is -.65 (see Figure 3, right panel). This value indicates how much the BMI gender gap among the middle educated differs from the gap among the middle educated, and exactly this is what we want to know. In a similar way the BMI gender gap for the highly educated is equal to: $b_1 + b_5$. The gender gap for the highly educated is -.83 while the gap among the low educated is .09 ($b_1$). So the difference between these two gaps ($b_5$) amounts to -.92 (-.83 – .09), see Figure 3 once more, right panel.

To summarize: when we like to test whether the BMI gender gap widens as the level of education rises, we have to look at the magnitude and p-value of the interaction effects $b_4$ and $b_5$. In Table 6 we present the results from our sample using the interaction equation (7):

**Table 6** The estimated BMI means for low education (a) and the mean differences with middle and high education for males ($b_2$ and $b_3$) and females ($b_2 + b_4$) and ($b_3 + b_4$)

|  | coefficients | standard error | t-value | p-value (2-tailed) |
|---|---|---|---|---|
| **Main effects** | | | | |
| a (constant) | 26.07 | .20 | 128.18 | .00 |
| female ($b_1$) | .09 | .28 | .31 | .75 |
| low | reference | | | |
| middle ($b_2$) | -.82 | .25 | -3.22 | .00 |
| high ($b_3$) | -1.37 | .26 | -5.37 | .00 |
| **Interaction effects** | | | | |
| female * low | reference | | | |
| female * middle ($b_4$) | -.65 | .35 | -1.86 | .06 |
| female * high ($b_5$) | -.92 | .36 | -2.57 | .01 |

In Table 6 some outcomes are identical to the results in Table 5: namely 26.07 (*a*), -.82 ($b_2$) and -1.37 ($b_3$). They relate to the mean BMI score for the low educated males (26.07), and the differences with the middle and highly educated males (-.82 and -1.37). If we add -.82 to 26.07, we get the estimated mean BMI for middle educated males (25.25) while 26.07 plus -1.37 equals the estimated mean BMI for high educated males (24.70).

The result for the variable *female* (.09, not significant) is the first BMI gender gap, when added to 26.07, we get the estimated mean BMI in low educated females (26.16). The BMI gender gap among the middle educated is -.56 (.09 plus the interaction effect -**.65**) and the third gap (high educated) is -.83 (.09 plus the interaction effect -**.92**). The mean BMI for the middle educated females is equal to 24.69 (26.07 + .09 + -.82 + **-.65**) and the mean BMI for high educated females is 23.87 (26.07 + .09 + -1.37 + **-.92**).

There are two equally valid ways to interpret the significant outcomes of the interaction effects $b_4$ (**-.65**) and $b_5$ (**-.92**). Firstly, the effect of education on BMI is stronger for females than for males. For males, the middle and high educated have BMI's that are -.82 and -1.37 below that of the low educated males (see Table 6). For females these differences are larger: -1.47 (-.82 + -**.65**) and -2.29 (-1.37 + -**.92**). This stronger effect of education for females is also visible in Figure 3, in which the downward BMI trend runs steepest for females.

The second interpretation is that the BMI gender gap is increasingly more negative as the educational level increases (the gap is .09 for low educated, -.56 for middle (.09 + **-.65**) and -.83 (.09 + **-.92**) for high educated). Both interpretations are equally true, and it depends on the research question which description is best suited. In this contribution we found it didactically useful to stress the BMI gender gap per educational level because it is more easily to understand compared to the focus on the unequal strength of the educational effect on BMI. It is remarkable that for the lower educated respondents a BMI gender gap hardly exists (the difference is .09 and not significant) and then reverses and grows wider for higher levels of education. In more general terms, this outcome –like in many other studies–, indicates that the BMI gender gap is not some biological constant but depends on other factors, like the educational level one has attained. We like to note that it is also possible to test whether the gender gap for the high educated is wider than that of the middle educated (difference in gender gap is -.27 (= -.83 − -.57)). To do this, one has to include a dummy for the low educated (and exclude the dummy for the middle educated) and include the interaction *female * low* (and exclude *female * middle*). Details can be found on our website.

## 7. Fifth example: more interactions, more variables

In our last example, we present the results from Table 4 once more, but this time we investigate whether next to education, the effect of having children is stronger for females than for males. In the previous example (Table 6) the results indicated that the effects of education were stronger in females (first interpretation of the interactions) and consequently that the BMI gender gaps widened with higher level of education (second interpretation of the interactions). In this fifth example we take things one last step further: we will test whether both the number of children and the level of education have a stronger impact on females' BMI, while controlling for year of interview and age. Because we have reasons to believe that the interaction effects are non-linear, we use interactions between the dummy variables. The equation used in this last example therefore is as follows:

> Mean (*BMI*) = a + $b_1$ * female + $b_2$ * middle + $b_3$ * high + $b_4$ * one child + $b_5$ * two children +
> $b_6$ * three children + $b_7$ * four or more children + $b_8$ * 2005 + $b_9$ * 2011
> + $b_{10}$ * age(log) + $b_{11}$ * female * middle + $b_{12}$ * female * high
> + $b_{13}$ * female * one child + $b_{14}$ * female * two children
> + $b_{15}$ * female * three children + $b_{16}$ * female * four or more children (8)
> - Reference categories: males, low education, no children, and the year 2000

At first glance Equation (8) looks complex, but it is rather straightforward after splitting into its components. The parameter *a* is the mean BMI for those respondents with code 0 on all variables included. In this case the mean BMI *a* relates to respondents who are: male, low educated, with no children, interviewed in 2000, and who are 18 years of age (we again use the re-scaled log-transformed age variable to have the 18 year-old respondents as a baseline). The parameters $b_1$ to $b_9$ all belong to dummy variables and show the controlled BMI difference with the related reference category. Because of the interactions, $b_2$ to $b_7$ relate to the mean BMI of males. For instance $b_2$ denotes the estimated mean BMI difference between middle and low educated males. The parameters $b_8$ and $b_9$ denote the difference between the mean BMI in 2000 (reference category) and 2005 respectively 2010, after controlling for all other variables. Note that we did not look for gender differences here: $b_8$ and $b_9$, are assumed to be equal for both males and females. The parameter $b_{10}$ for the variable *age*(log) indicates the effect of being older on the mean BMI, again after taking account the other factors and again without considering the possible different age effects among males and females (see our third example and our webpage for more details about this age effect).

The parameters $b_{11}$ to $b_{16}$ are of most interest here because they show the extent to which BMI effects of education and number of children are different for females. We mentioned earlier that the $b_2$ parameter tells us how much the mean BMI in middle educated males differ from low educated males. The $b_{11}$ parameter indicates how much more/less this difference is between low and middle educated females.

In Table 7 we show the results for Equation (8). We will go through all parameters once more to summarize all that we addressed about the use of dummy variables and their interactions. The parameter *a* (also known as 'constant' or 'intercept') amounts to 23.36. This is the mean BMI for all the respondents with score 0 on all variables included. These people are therefore male, low educated with no children, they were interviewed in the year 2000 and are 18 years old (age centered on age 18).

**Table 7** A multiple regression model for BMI including sex, education, year, number of children, and age with interactions for education and number of children

|  | coefficients | standard error | t-value | p-value (2-tailed) |
|---|---|---|---|---|
| **Main effects** |  |  |  |  |
| a (constant) | 23.36 | .30 | 78.79 | .00 |
| female ($b_1$) | .14 | .36 | .39 | .69 |
| low | reference |  |  |  |
| middle ($b_2$) | -.45 | .25 | -1.79 | .07 |
| high ($b_3$) | -1.07 | .25 | -4.27 | .00 |
| no children | reference |  |  |  |
| 1 child ($b_4$) | 1.20 | .32 | 3.74 | .00 |
| 2 children ($b_5$) | .78 | .24 | 3.20 | .00 |
| 3 children ($b_6$) | 1.03 | .31 | 3.31 | .00 |
| 4 children or more ($b_7$) | 1.09 | .43 | 2.56 | .01 |
| 2000 | reference |  |  |  |
| 2005 ($b_8$) | .21 | .15 | 1.36 | .17 |
| 2011 ($b_9$) | .39 | .17 | 2.38 | .02 |
| age (log) ($b_{10}$) | 1.92 | .22 | 8.57 | .00 |
| **Interaction effects** |  |  |  |  |
| female * low | reference |  |  |  |
| female * middle ($b_{11}$) | -.52 | .34 | -1.51 | .13 |
| female * high ($b_{12}$) | -.72 | .35 | -2.03 | .04 |
| female * no children | reference |  |  |  |
| female * 1 child ($b_{13}$) | -.45 | .42 | -1.06 | .29 |
| female * 2 children ($b_{14}$) | -.29 | .32 | -.91 | .36 |
| female * 3 children ($b_{15}$) | -.39 | .40 | -.98 | .33 |
| female * 4 children or more ($b_{16}$) | -.40 | .58 | -.68 | .50 |

The mean BMI among females is .14 points higher than among males. This small and not significant gender gap is conditional however: it applies to respondents with a low education who have no children, because of the interactions parameters included for *education* (middle and high) and *number of children* (one, two, three and four or more children).

The mean BMI of the middle educated males lies -.45 below that of low educated males. The BMI among highly educated males is on average -1.07 points lower compared to the lower educated males. We further find that having children does increase a male's BMI: males with one, two, three, and four or more children have on average a BMI that is about 1 point higher compared to males with no children. For all respondents, regardless whether they are female or male, the mean BMI increases over the years: in 2005 the mean BMI (after controlling for all other variables) is .21 (not significant) higher than in 2000. In 2011 the mean BMI is .39 higher than in 2000. Also the age effect is significant and curve-linear: as people (both male and female) are older, the BMI increases but this effect gradually decreases. As a short note we like to add that it is difficult to identify the 'real' age effect (i.e., the effect of growing older) in our models, because we did not follow our respondents over time as they actually grow older, and because we investigate only a very limited period (2000-2011). In fact, we just observe that older people in our sample have a higher BMI than younger people and we cannot tell how much of this effect is related to the biological process of growing older.

Next in Table 7 are the outcomes for the interactions. Generally, interaction parameters indicate that effects vary across groups/categories. We already found in our fourth example that the mean BMI across the tree levels of education for males is not as differentiated as for females (see Figure 3). In this example we also tested whether having children has a stronger impact on females' BMI.

Table 7 shows that the difference between middle educated females and low educated females is not -.45 (the difference between low and middle educated males) but an extra -.52 points lower (this extra difference of -.52 is significant in a one-tailed test with α=.10). So, the total BMI difference between middle and low educated females is -.97 (-.52 + -.45). Also, the difference between low and high educated females is larger than for males (the extra difference amounts to a significant -.72 (the total difference is -1.07 plus -.72). So, the negative effect of education (the higher education, the lower BMI) again is found to be stronger in females. This is the first interpretation of this interaction between education and sex. In terms of the BMI gender gap (the second and equally true interpretation), the interaction tells us that the initial gender gap among the lower educated widens as education is higher. For the middle educated the gap is -.52 points wider and for the high educated it is even -.72 points wider. We like to note that these two interaction effects ($b_{11}$ and $b_{12}$) show the difference with the gender gap among the low educated regardless how many children one has.

We also tested whether having children has a significant larger impact on females' BMI compared to males. This turns out not to be the case as all relevant interactions have a negative sign. This means that the difference in mean BMI between females with and without children is somewhat smaller compared to the differences among males. For example, the controlled difference between males with no children (the reference category) and males having one child is +1.20 points, whereas the difference between childless females and females with one child is: 1.20 + -0.45 = 0.75. Note that we hypothesized that having children would be more detrimental to the BMI of females, but oddly it tends to be the other way around. However, we cannot claim with enough statistical confidence that the impact of having children is larger for males as all interactions are not significant by reasonable standards.

If one wants to visualize the BMI gender gaps in this last example in a graph then we can do this a) for the 3 educational levels, b) for the 5 categories of the number of children, and c) for the 3 * 5 combinations of these categories. We plotted these gaps on our website, just to show how education and the number of children *separately* and *combined* interact with (or moderate) the BMI gender gap.

## 8. Conclusions

In regression analysis the use of 'dummy' variables is common practice. In this contribution we discussed a very popular type, namely *dummy coding*. It is characterized by its use of a reference category. All estimates of the related dummy variables are deviations from that particular reference, eventually after taking into account other variables. This is particular useful if one wants to test directional hypotheses, for example the notion that the mean BMI tends to be lower as someone's educational level is higher. If one is more agnostic about the direction of effects, then dummy coding is less suited. Instead researchers may then use 'effect coding' (also called ANOVA coding) in which the reference point shifts to the unweighted mean. This is especially useful in experimental settings where group sizes are irrelevant. An alternative is 'weighted effect coding' in which the reference is the sample mean, and group sizes are thus taken into account.

On our website we provide all data and syntax to obtain all results discussed here. We also added some text and syntax to test interactions between dummy variables and interval/ratio variables and interactions between two interval/ratio variables together with graphs. We finally present the use of effect coding and weighted effect coding and their interactions.

## References

**Eisinga** Rob, Coenders Marcel, Felling Albert, te Grotenhuis Manfred, Oomens Shirley, Scheepers Peer. (2002) Religion in Dutch society 2000. Documentation of a national survey on religious and secular attitudes in 2000. NIWI-Steinmetz Archive, Amsterdam. http://www.ru.nl/sociology/research/socio-cultural/

**Eisinga** Rob, Need Arianne, Coenders Marcel, de Graaf Nan Dirk, Lubbers Marcel, Scheepers Peer, Levels Mark, Thijs, Paula. (2012a) Religion in Dutch society 2005. Documentation of a national survey on religious and secular attitudes and behaviour in 2005, DANS Data Guide 10. Amsterdam University Press, The Hague. http://www.ru.nl/sociology/research/socio-cultural/

**Eisinga** Rob, Kraaykamp Gerbert, Scheepers Peer, Thijs Paula. (2012b) Religion in Dutch society 2011-2012. Documentation of a national survey on religious and secular attitudes and behaviour in 2011-2012, DANS Data Guide 11. Amsterdam University Press, The Hague. http://www.ru.nl/sociology/research/socio-cultural/

**Hardy**, Melissa. (1993). Regression with dummy variables. Los Angeles: Sage, little green book series: http://srmo.sagepub.com/view/regression-with-dummy-variables/n1.xml

**Hermann**, Silke, Rohrmann, Sabine, Linseisen, Jacob, ... & Peeters, Petra. (2011). The association of education with body mass index and waist circumference in the EPIC-PANACEA study. *BMC Public Health*,*11*(1), 169. http://www.ncbi.nlm.nih.gov/pubmed/21414225

**Krul** Arno J., Hein A. M. Daanen and Hyegjoo Choi (2011). Self-reported and measured weight, height and body mass index (BMI) in Italy, the Netherlands and North America. *European Journal of Public Health, 21*(4), 414-419. http://eurpub.oxfordjournals.org/content/21/4/414

**Stevens**, Gretchen, Gitanjali M Singh, Yuan Lu,... and Majid Ezzati (2012). National, regional, and global trends in adult overweight and obesity prevalences. *Population Health Metrics*, 10:22. http://www.pophealthmetrics.com/content/10/1/22

**Umberson**, Debra, Lui H., Mirowsky, J. & Reczek, Corinne. (2011). Parenthood and trajectories of change in body weight over the life course, *Social Science and Medicine, 73*(9), 1323-1331. http://www.ncbi.nlm.nih.gov/pmc/articles/PMC3391503/